\documentstyle[aas2pp4]{article}
\def \mbh {$M_{\rm BH}$(H$\beta$)}
\begin{document}
\title{ON BLACK HOLE MASSES AND RADIO LOUDNESS IN AGN}
\author{Ari Laor}
\affil{Physics Department, Technion, Haifa 32000, Israel\\
laor@physics.technion.ac.il}

\begin{abstract}
The distribution of radio to optical fluxes in AGN is
bimodal. The physical origin for this bimodality is not understood.
In this {\em Letter} I describe observational evidence, based on
the Boroson \& Green PG quasar sample, that the radio loudness 
bimodality is strongly related to the black 
hole mass ($M_{\rm BH}$). Nearly all PG quasars with $M_{\rm BH}>10^9M_{\odot}$ 
are radio loud, while quasars with $M_{\rm BH}<3\times 10^8M_{\odot}$ are 
practically all radio quiet. This result is consistent with the dependence of
quasar host galaxy morphology on radio loudness. There is no simple physical 
explanation for this result, but it may provide a clue on how jets are formed 
near massive black holes. The 
radio loudness--black hole mass relationship suggests that the properties of various 
types of AGN may be largely set by three basic parameters, $M_{\rm BH}$, 
$L/L_{\rm Eddington}$, and inclination angle.

\end{abstract}


\section{INTRODUCTION}
The radio to optical flux distribution in AGN is bimodal. This is demonstrated
most clearly in the recent large compilation of Xu, Livio, \& Baum (1999),
which shows that radio loud AGN are about $10^4$ times brighter in the radio 
than radio quiet AGN with the same [O~III] luminosity (which provides a measure 
of the ionizing continuum luminosity). The reason for this bimodality is one of 
the basic unsolved problems in AGN physics (e.g. Krolik 1999, Ch.15).

The radio emission is produced by relativistic electrons which are powered
by a jet, both in radio loud AGN (e.g. Begelman, Blandford, \& Rees 1984), 
and apparently also in radio quiet AGN (Blundell \& Beasley 1998). What then 
controls the jet power, and why is the
relative (i.e. radio to bolometric) power distribution bimodal?  
An ion torus may be required to collimate
the jet, a spinning black hole may be required to accelerate the jet, 
and a low density environment may be required to allow it to propagate
(e.g. Blandford \& Znajek 1977; Blandford \& Levinson 1995; Fabian \& Rees 1995; 
Moderski, Sikora, \& Lasota 1998; Rees et al. 1982; Wilson \& Colbert 1995). 
However, no strong observational evidence is currently available
to support these, or any other scenarios. 
  
Hubble Space Telescope (HST) observations over the past few years 
established that radio loud quasars
always reside in bright elliptical (or sometimes interacting) hosts, and that 
all quasars with spiral hosts are radio quiet (e.g. Bahcall et al. 1997; 
McLure et al. 1999). This relation is
puzzling, {\em how does the inner mpc of a galaxy, where the
jet originates, knows about the type of host it resides in?}

Some recent observations, and the new evidence described in this
paper, provide a clue for one of the basic parameters which appears to
control the 
formation of powerful jets, as further described below (see Laor 2000 for
a short account).

\section{EXISTING EVIDENCE}

Xu, Livio, \& Baum (1999) noted that radio loud AGN extend to higher
[O~III] luminosity than radio quiet AGN, and suggested that this may imply 
that the
distribution of black hole masses in radio loud AGN extends to higher
masses. Corbin (1997) made a similar suggestion based on the tendency
of radio loud AGN to have broader H$\beta$ lines.

A number of studies over the past few years established a few
correlations which point more directly towards a relation between
radio loudness and black hole mass, as further described below.

Compact non-thermal radio emission is commonly detected in the nuclei
of normal elliptical galaxies (e.g. Sadler, Jenkins \& Kotanyi 1989),
and also in some spiral galaxies (Sadler et al. 1995).
Similar emission is common in Seyfert galaxies 
(e.g. Nelson \& Whittle 1996),
and obviously in radio galaxies as well, where it is correlated with 
the nuclear H$\alpha$ emission (e.g. Zirbel \& Baum 1995). 
Ho (1999a) has shown that the nuclear radio power $L_R$ vs. H$\alpha$ 
luminosity correlation
extends down to the lowest powers observed in nearby ellipticals, 
suggesting that their radio emission originates in a 
scaled down AGN.

Nelson \& Whittle (1996) explored relations between the bulge properties
and the AGN properties in a large sample of Seyfert galaxies.
They found a correlation between $L_R$ and the bulge
luminosity and velocity dispersion, which implies a correlation between
$L_R$ and the bulge mass (as suggested by Heckman 1983).

Magorrian et al. (1998), studied the 
demography of massive black holes in nearby galaxies, and found that possibly all
bulge galaxies have a massive black hole with a mass which correlates with 
the bulge mass, as first suspected by 
Kormendy (1993). This correlation is further supported by the recent studies 
of Gebhardt et al. 
(2000) and Ferrarese \& Merritt (2000).

If the radio power is correlated with the bulge mass, 
and the bulge mass is correlated with the black hole mass
(in both active and non active galaxies), then the radio
power may be directly linked with the black hole mass. Indeed, Franceschini,
Vercellone \& Fabian (1998) found a surprisingly tight relation between 
black hole mass and radio power in a small sample of nearby mostly non active
galaxies. McLure et al. measured the host properties of a sample of
AGN, made an indirect estimate of their $M_{\rm BH}$ using the Magorrian et al. 
relation, and found their objects follow the $M_{\rm BH}$ vs. $L_R$ relation of 
Franceschini et al.

Thus, it is interesting to explore whether there is a direct 
relation between $M_{\rm BH}$ and $L_R$ in AGN as found for nearby galaxies, 
and also whether the radio 
loudness bimodality is related in any way to $M_{\rm BH}$.

\section{THE NEW EVIDENCE}

In order to explore the $M_{\rm BH}$ vs. $L_R$ relation directly in AGN ones needs
a direct way to estimate $M_{\rm BH}$. A long known method to deduce $M_{\rm BH}$
is to use the broad emission line width, and the distance of the Broad Line 
Region (BLR) from the center, 
together with the assumption of Keplerian motion (e.g. Dibai 1980). 
This method was subject to 
unknown, but potentially large errors, due to unestablished
assumptions concerning the BLR radius, dynamics, and kinematics.  
Significant progress in reverberation
mappings over the past few years established the radius luminosity relation
for the BLR (Kaspi et al. 2000), and strongly suggests Keplerian dynamics
in a few well explored cases (e.g. Peterson \& Wandel 2000). 
However, possible anisotropy of the ionizing
continuum, and of the cloud kinematics, still leaves a room for potentially 
significant systematic errors. 

The H$\beta$ line width and continuum luminosity 
were used by Laor (1998) to derive \mbh\ for a sample of Palomar Green (PG) 
quasars (Schmidt \& Green 1983)
observed by Bahcall et al. with the HST. 
This study revealed that 
\mbh\ is correlated with the bulge luminosity, 
and that this correlation overlaps remarkably well the Magorrian et al.
correlation. This overlap provides an indirect check for the accuracy of the 
\mbh\ estimate, and
indicates that any remaining systematic errors are less than a factor of 
$2-3$ (Laor 1998). This check is particularly important for the radio loud AGN, 
where the generally large width of H$\beta$ could otherwise be attributed to jet 
interactions
with the BLR, as was suggested by Whittle (1992) for the forbidden lines
of radio loud AGN. 

To explore the $M_{\rm BH}$ vs. $L_R$ relation in quasars I use the
Boroson \& Green  (1992) sample of all 87 $z<0.5$ PG quasars, where
they provide H$\beta$ FWHM values 
based on their high quality optical spectra.
\footnote{with the following corrections: PG 1307+085 FWHM=5320~km~s$^{-1}$,
PG 2304+042 FWHM=6500~km~s$^{-1}$}
The optical continuum luminosity is taken from Neugebauer et al. (1987).
These parameters are combined, as in Laor (1998), to yield
$m_9=0.18\Delta v_{3000}^2 L_{46}^{1/2}$,
where $m_9\equiv M_{\rm BH}({\rm H}\beta)/10^9M_{\odot}$, \\ 
 $\Delta v_{3000}\equiv {\rm H}\beta$~FWHM/3000~km~s$^{-1}$, and\\ 
$L_{46}=L_{\rm bol}/10^{46}$~erg~s$^{-1}$, where the bolometric
luminosity is
$L_{\rm bol}=8.3\times \nu L_{\nu}$(3000\AA). Kaspi et al. suggest
a somewhat steeper radius luminosity relation for the BLR 
than assumed above, but this has a small effect ($<50$\%) 
on the mass estimates of most objects. The radio luminosity
$L_R\equiv \nu L_{\nu}$(5~GHz) is obtained from Kellermann et al. (1989),
modified for $H_0=80$~km~s$^{-1}$~Mpc$^{-1}$, and $\Omega_0=1$ adopted here.

Franceschini et al. found a tight relationship between 
$M_{\rm BH}$ and $L_R$ based on a compilation of these
parameters for 13 nearby weakly 
or non active galaxies. A larger 
sample of 29 nearby galaxies is obtained here by combining 
all $M_{\rm BH}$ values from Magorrian et al., and 
Gebhardt et al. (which supersedes some of the Magorrian et al. values),
with all single dish 5~GHz $L_R$ values from 
Fabbiano, Gioia \& Trinchieri (1989), and Becker, White, \& 
Edwards (1991).

Figure 1 shows the $M_{\rm BH}$ vs. $L_R$ relation for the 87 PG quasars, 
and the 29 nearby galaxies, together with the linear relation found by
Franceschini et al. The scatter is very large. Nearby galaxies display
a range of typically $10^4$ in $L_R$ at a given $M_{\rm BH}$, and this range
increases to $10^6$, or more, when active galaxies are included.

There is certainly a trend of $L_R$ increasing with $M_{\rm BH}$,
but the tight relation suggested by Franceschini et al.  
is not supported by our data. This trend is more apparent
for the quasars. In particular, there appears to be a rather sharply
defined "zone of avoidance", where the maximum radio luminosity 
$L_R^{\rm max}$, at
a given $M_{\rm BH}$,
increases with $M_{\rm BH}$. The increase in $L_R^{\rm max}$ is highly 
nonlinear, going up from $\sim 5\times 10^{38}$~erg~s$^{-1}$ for 
$10^7M_{\odot}$, to $\sim 5\times 10^{40}$~erg~s$^{-1}$ for 
$10^8M_{\odot}$, to $>10^{44}$~erg~s$^{-1}$ for 
$10^9M_{\odot}$. 

The very large range of $L_R$ at a given $M_{\rm BH}$ is not surprising, 
it may simply be due to different levels of overall continuum luminosity 
of different AGN with the same black hole mass. However, how is the 
{\em fraction} of the bolometric luminosity emitted in the radio
dependent on $M_{\rm BH}$?

Figure 2 shows the relation between $M_{\rm BH}$ and the radio loudness
parameter $R\equiv f_{\nu}({\rm 5 GHz})/f_{\nu}({\rm 4400\AA})$ 
for the Boroson \& Green sample, as taken from Kellermann et al. 
The distribution of $R$ values
is bimodal, with a minimum at $R=10$, commonly used to define radio 
loud vs. radio quiet quasars. 
The distribution of $M_{\rm BH}$ for the radio loud and radio quiet
PG quasars is remarkably different. 
Most quasars (10/11) with $M_{\rm BH}>10^9M_{\odot}$ are radio loud,
and essentially all quasars with $M_{\rm BH}<3\times 10^8M_{\odot}$
are radio quiet. The probability that the radio loud and radio quiet 
PG quasars are drawn
from the same mass distribution is $4\times 10^{-7}$ according to 
the KS test (using the KSTWO routine of Press et al. 1992). 
Interestingly, despite the highly significant difference in mass
distribution, the difference in the distribution of $L/L_{\rm Eddington}$
values ($\propto L_{\rm bol}/m_9$) is much less significant
($4.4\times 10^{-2}$).

\section{DISCUSSION}

The tight relation between radio luminosity and black hole mass 
suggested by Franceschini et al. is not supported by our larger sample 
of 29 nearby galaxies. The scatter becomes even larger when active
galaxies are included. For example, at $M_{\rm BH}=3\times 10^8M_{\odot}$ the
radio power ranges from $<10^{35}$~erg~s$^{-1}$ to
$10^{42}$~erg~s$^{-1}$ (Fig.1). 
The non-thermal emission associated with a massive
black hole is high when the system is active, but it can be very weak to non 
detectable when the system is generally inactive.

However, the radio/optical luminosity ratio, or equivalently
the {\em relative} jet power, in active galaxies
is strongly related to the black hole mass (Fig.2). A high mass
black hole, $M_{\rm BH}>3\times 10^8M_{\odot}$, is necessary for
a relatively powerful jet, and is sufficient if
$M_{\rm BH}>10^9M_{\odot}$. Conversely, relatively powerful jets are
impossible if $M_{\rm BH}<3\times 10^8M_{\odot}$. {\em Why should
the relative jet power be so critically
dependent on $M_{\rm BH}$?}  

None of the models for the formation of powerful jets (mentioned in 
\S 1) predicts such a strong dependence on $M_{\rm BH}$. Some of these
models suggest that radio loud AGN should have a low $L/L_{\rm Eddington}$
(e.g. Rees et al.), but as mentioned in \S 3, 
the observed dependence on $M_{\rm BH}$
is much stronger than the dependence on $L/L_{\rm Eddington}$.

One physical process which is directly linked to $M_{\rm BH}$ is tidal 
disruption of main sequence stars outside the event horizon, which 
ceases for $M_{\rm BH}>2\times 10^8M_{\odot}$ for a rotating black hole 
(e.g. Rees 1988). However, tidal
disruption is likely to be rather intermittent ($<10^{-1}$~yr$^{-1}$, Rees), 
and it 
is not clear why its effects (e.g. the disruption of a jet maintaining
B field configuration) should be so long lasting, compared to the
local dynamic timescale ($<$day).
Alternatively, if jets are powered by the black hole spin, then the above 
correlation may result from a tight relation between black hole mass and 
spin. 

Falcke, Sherwood, \& Patnaik (1996) cautioned that since radio quiet quasars
may also be powered by jets, some of the apparently radio loud quasars
in the PG sample 
could be intrinsically radio quiet quasars which are beamed at us. These
Radio Intermediate Quasars (RIQ)
can be identified through their flat radio spectra ($\alpha>-0.5$, 
indicating
core dominance), yet relatively low $R$ values ($\le 250$) compared to flat 
spectrum radio selected quasars. There are four RIQ in the Boroson \& Green 
sample,
PG~0007+106, PG~1302$-$102, PG~1309+355, and PG~2209+184.
VLBI observations of three of these confirmed their 
highly compact sizes (mas), as expected under the beaming hypothesis 
(Falcke, Patnaik, \& Sherwood 1996), and also revealed superluminal motion in
one (PG~0007+106, Brunthaler et al. 2000). These four RIQ are marked
in Fig.2. It is interesting that these
four objects all fall at the lowest $M_{\rm BH}$ values of the radio loud
quasars. If these quasars are indeed all intrinsically radio quiet, then 
radio loud and radio quiet AGN may overlap only in the range
$5\times 10^8M_{\odot}-10^9M_{\odot}$, and given the likely
uncertainty in the $M_{\rm BH}$ estimate, may not overlap at all.

The $M_{\rm BH}$ vs. bulge mass relation, together with the 
$R$--$M_{\rm BH}$ relationship, provides a phenomenological understanding
of the relation between radio loudness and host properties.
Spiral galaxies have small bulges, these bulges have 
low mass black holes, and these cannot produce radio loud AGN. A radio loud 
quasar requires a massive black hole, and this is found only in bright ellipticals. 
Elliptical galaxies can have a low luminosity, thus 
a black hole mass below $10^9M_{\odot}$, and
thus host radio quiet AGN, as observed.
Similarly, BL Lac objects, which are always radio loud, are essentially always 
found in luminous elliptical hosts (e.g. Urry et al. 2000).

Lacy, Ridgway \& Trentham (2000) have noted that the Magorrian et al. 
relation, together with the fact that radio loud AGN reside in bright ellipticals,
``strongly suggests a link between radio loudness and black hole mass'', and
further proposed this can explain the increase in the fraction of radio loud quasars 
with luminosity, from $<10$\% at M$_{\rm B}>-24$ to $\sim 50$\% at
M$_{\rm B}=-28$ seen in some surveys (Hooper et al. 1996; Goldschmidt et al. 
1999).
This rise is consistent with the $R$--$M_{\rm BH}$ relationship since a 
magnitude of M$_{\rm B}=-28$ corresponds to 
$\nu L_{\nu}$(4400\AA)$\sim 5\times 10^{46}$~erg~s$^{-1}$, or
$L_{\rm bol}\sim 5\times 10^{47}$~erg~s$^{-1}$, and thus
if the Eddington limit applies, then $M_{\rm BH}>4\times 10^9M_{\odot}$.
However, Stern et al.
(2000) find the fraction of $z>4$ radio loud quasars to be constant up to
M$_{\rm B}=-28$. Thus, the validity of the $R$--$M_{\rm BH}$ relationship
at high redshifts remains an open question.

The $R$--$M_{\rm BH}$ relationship may help explain the low fraction 
of radio loud AGN at M$_{\rm B}>-24$ in some quasar surveys (e.g. Hooper et 
al.). 
Radio loud AGN necessarily reside in bright hosts, and if the AGN is
weak the object will be classified as a ``radio galaxy'', and
may be rejected from optical quasar surveys 
due to color or morphology criteria. 
Radio quiet AGN can reside in fainter hosts, and thus be 
easier to detect in quasar surveys down to lower luminosity.

How far down in luminosity is the $R$--$M_{\rm BH}$ relationship maintained?
The relation between bulge luminosity and $L_R$ presented by Nelson \& Whittle
(1996) suggests (through the
$M_{\rm BH}$ vs. $M_{\rm bulge}$ relation) that the $R$--$M_{\rm BH}$ relationship
holds down to the Seyfert luminosity level. Further down, at the 
very weakly active galaxies level little data is currently available. 
Ho (1999b) provides a rough spectral energy distribution for 
seven very weak AGN 
($L_{\rm bol}\sim 10^{41}-10^{42}$~erg~s$^{-1}$)
with measured $M_{\rm BH}$.
The standard $R$ parameter may not be a
useful indicator of the relative jet power in these AGN 
since the optical emission
carries a very small fraction of $L_{\rm bol}$. I therefore use
$L_R/L_{\rm bol}$ instead of $R$ for the relative jet power, where
$L_R$ is obtained from single dish
broad beam (rather than VLBI) radio fluxes, to roughly match the spatial 
scales measured for the PG quasars. 
The two AGN in the sample of Ho with $M_{\rm BH}=4\times 10^6 M_{\odot}$ have
$\langle \log L_R/L_{\rm bol}\rangle = -4.1$, while the other
five AGN with $M_{\rm BH}\ge 5\times 10^8 M_{\odot}$ 
have $\langle \log L_R/L_{\rm bol}\rangle = -2.2$. 
This suggests that the $R$--$M_{\rm BH}$ relationship extends 
down to very low AGN activity levels.
Interestingly, the jets in the Galactic microquasars, which most likely harbor
$\sim 10 M_{\odot}$ black holes, are also ``radio quiet'' 
(e.g. Mirabel \& Rodriguez 1998, Log $L_R/L_{\rm bol}\sim -7$).

How sharp is the transition in $R$ with $M_{\rm BH}$?  
The $R$--$M_{\rm BH}$ relationship is established
here only for $z\le 0.5$ optically selected bright AGN. It is important to
study this relationship in a similarly complete, well 
defined, and deep sample of radio selected AGN, such as the FIRST Bright 
Quasar Survey sample (although this sample includes relatively few 
``proper'' radio quiet AGN). Optical spectroscopy of a relatively 
large and
heterogeneous sample of radio selected quasars is presented by 
Brotherton (1996). The lack of accurate spectrophotometry, non uniform 
spectroscopy, and sample inhomogeneity do not allow
one to draw robust conclusions on the  
$R$--$M_{\rm BH}$ relationship. However, at the order of magnitude
level, one finds that all the newly measured quasars
in this sample (except one, 3C~232) appear to have $M_{\rm BH}\ge 
10^8M_{\odot}$, and $\sim 2/3$ appear to have 
$M_{\rm BH}>3\times 10^8M_{\odot}$. 

If radio loudness is indeed set by $M_{\rm BH}$, then it may be
possible to relate the various types of AGN to various combinations of
just three basic parameters, $M_{\rm BH}$, $L/L_{\rm Edd}$, and the 
inclination angle $\theta$.
Figure 3 provides a rough sketch of the likely positions of the various
types of AGN in the $M_{\rm BH}$, $L/L_{\rm Edd}$, $\theta$ cube.
All radio loud AGN are located on the high $M_{\rm BH}$ side,
and all AGN where the bulge light is significant, or dominant, are
necessarily on the low $L/L_{\rm Edd}$ side. The position along the 
$\theta$ axis is derived from inclination based unification schemes 
which are now quite well established (Antonucci 1993; Urry \& Padovani 1995; 
Wills 1999 ).

\acknowledgments

I thank the referee for a knowledgeable and helpful report.
This research was supported by the fund for the promotion of research
at the Technion.

\onecolumn

\newpage
\begin{figure}
\plotone{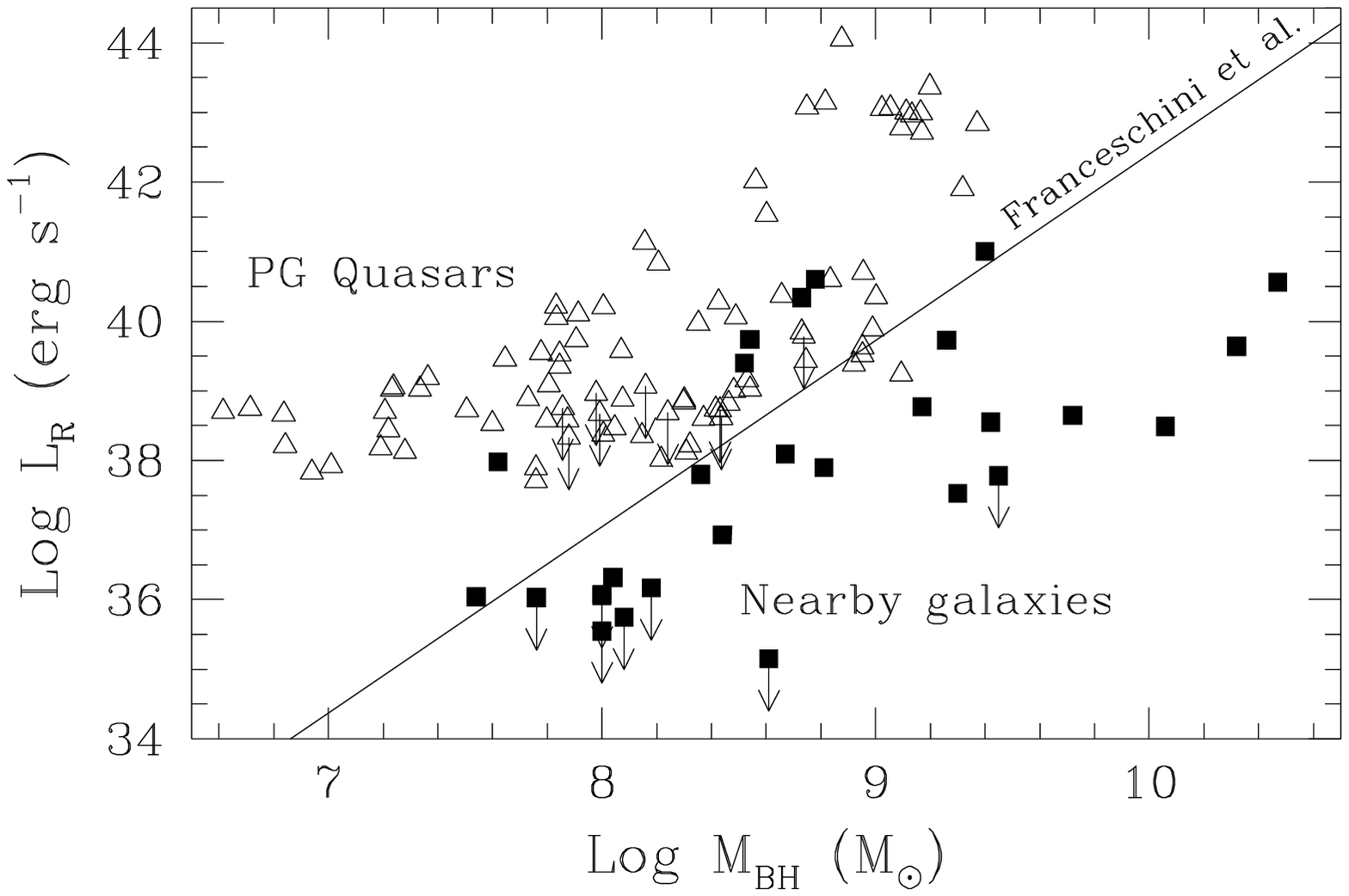}
\caption{The dependence of radio luminosity on black hole mass. Nearby normal
galaxies are marked as filled squares, and the PG quasars as empty triangles.
Points with downward arrows indicate upper limits. The tight relation found
by Franceschini et al. for a smaller sample of nearby galaxies is indicated
by the solid line. Note the very large scatter in this relation
for both active and non active galaxies.}
\end{figure}

\begin{figure}
\plotone{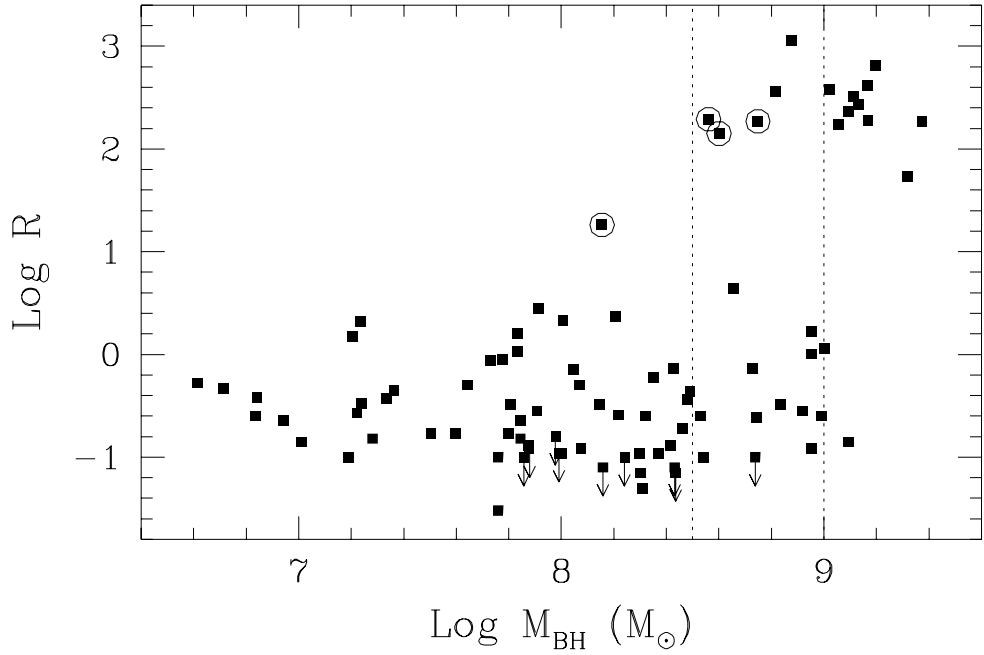}
\caption{The radio loudness versus black hole mass for the Boroson \& Green sample
of 87 $z\le 0.5$ PG quasars. Most quasars with $M_{\rm BH}>10^9M_{\odot}$ are radio 
loud (Log~$R>1$), and essentially all quasars with $M_{\rm BH}<3\times 10^8M_{\odot}$
are radio quiet. The objects surrounded by a circle were proposed by Falcke et al.
to be beamed intrinsically radio quiet quasars (see text), if true then radio loud 
and radio quiet quasars may not overlap in $M_{\rm BH}$}.
\end{figure}

\begin{figure}
\plotone{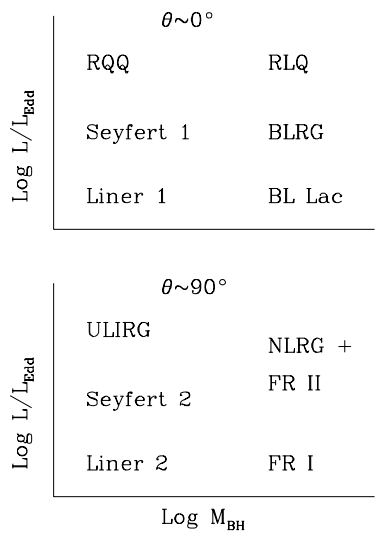}
\caption{A schematic representation of an $M_{\rm BH}$, $L/L_{\rm Edd}$, 
$\theta$ unification scheme. All objects at high $M_{\rm BH}$ are radio loud. 
All objects at low $L/L_{\rm Edd}$
have a low AGN/bulge luminosity ratio. All objects at $\theta\sim 90^{\circ}$ 
have an obscured core (the standard AGN unification scheme).}
\end{figure}

\end{document}